\documentclass[superscriptaddress,twoside,twocolumn,bibnotes,tightenlines,aps,floats]{revtex4}
\usepackage{times}
\usepackage{graphicx}

\begin{document}
\graphicspath{{../Figures/}{.}}
\title{Four-phase patterns in forced oscillatory systems} 
\author{A.~L. Lin}
\affiliation{Center for Nonlinear Dynamics and Department of Physics,
The University of Texas at Austin, Austin, TX 78712}
\author{A. Hagberg} \email{aric@lanl.gov}
\affiliation{Center for Nonlinear Studies and T-7, Theoretical Division,
Los Alamos National Laboratory, Los Alamos, NM 87545}
\author{A. Ardelea}
\affiliation{CFD Laboratory and ASE/ME Department,
The University of Texas at Austin, Austin, TX 78712}
\author{M. Bertram}
\author{H.~L. Swinney}
\affiliation{Center for Nonlinear Dynamics and Department of Physics,
The University of Texas at Austin, Austin, TX 78712}

\author{E. Meron} \email{ehud@bgumail.bgu.ac.il}
\affiliation{Department of Energy and Environmental Physics, BIDR, 
and Physics Department, Ben-Gurion University, Sede Boker Campus 84990, Israel}

\begin{abstract}
We investigate pattern formation in self-oscillating systems forced by
an external periodic perturbation.  Experimental observations and
numerical studies of reaction-diffusion systems and an analysis of an
amplitude equation are presented. The oscillations in each of these
systems entrain to rational multiples of the perturbation frequency
for certain values of the forcing frequency and amplitude.  We focus
on the subharmonic resonant case where the system locks at one fourth
the driving frequency, and four-phase rotating spiral patterns are
observed at low forcing amplitudes.  The spiral patterns are 
studied using an amplitude equation for periodically forced
oscillating systems.  The analysis predicts a bifurcation (with
increasing forcing) from rotating four-phase spirals to standing
two-phase patterns.  This bifurcation is also found in periodically
forced reaction-diffusion equations, the FitzHugh-Nagumo and
Brusselator models, even far from the onset of oscillations where the
amplitude equation analysis is not strictly valid.  In a
Belousov-Zhabotinsky chemical system periodically forced with light we
also observe four-phase rotating spiral wave patterns.  However, we
have not observed the transition to standing two-phase patterns,
possibly because with increasing light intensity the reaction kinetics
become excitable rather than oscillatory.
 
\end{abstract}

\received{23 February 2000}

\maketitle

\section{Introduction}

Spatially extended systems characterized by the coexistence of two or
more stable states compose a broad class of nonequilibrium pattern
forming systems.  The most common multistable systems are those that
exhibit bistability, (e.g.  chemical systems~\cite{KaSh:94,POS:97},
vertically vibrated granular systems~\cite{UMS:96}, and
binary fluid convection~\cite{KBS:88}).  
Spatial patterns in these systems involve
alternating domains of the two different stable states, which are
separated from each other by interfaces or fronts.  Bistable systems
support a variety of patterns from spiral waves to splitting spots and
labyrinths~\cite{Pear:93,LeSw:95,HaMe:94b,HaMe:94c,GMP:96}.  In some
systems, such as the ferrocyanide-iodate-sulfite
reaction~\cite{LeSw:95,LMOS:93} and the oxidation of carbon monoxide
on a platinum surface~\cite{HBKR:95}, the bistability arises from the
nonlinear nature of the system.  In other systems such as liquid
crystals in a rotating magnetic field \cite{MiMe:94,FRCG:94,FrGi:95}
and periodically forced oscillators~\cite{CoEm:92}, the bistability
arises from a broken symmetry.

Periodically forced oscillatory systems are convenient systems for
exploring multistability in pattern formation since the number of
coexisting stable states can be controlled by changing the forcing
frequency.  Applying a periodic force of sufficient amplitude and at a
frequency $\omega_f \approx \frac{n}{m}\omega_0$, where $\omega_0$ is
the oscillation frequency of the unforced system, entrains the system
to the forcing frequency.  The entrained system has $n$ stable states
each with the same oscillation frequency but in one of $n$ oscillation
phases separated by multiples of $2\pi/n$.  We refer to the $n$
different phase shifted states as ``phase states'' of the system.

Recent experiments using the ruthenium-catalyzed Belousov-Zhabotinsky
reaction forced by periodic illumination revealed subharmonic
resonance regimes $\omega_f : \omega_0 =$ 2:1, 3:2, 3:1, 4:1, with two
(2:1) , three (3:2,3:1), and four (4:1) stable phase
states~\cite{POS:97,LBMS:99}.  Patterns consisting of alternating
spatial domains with a phase shift of $\pi$ are observed within the
2:1 resonance regime, and three-phase patterns with spatial domains
phase-shifted by $2\pi/3$ are observed within the 3:1 resonance
regime~\cite{POS:97,LBMS:99,LPACW:99}.  The 4:1 resonance is more
complicated.  Adjacent spatial domains may differ in phase by either
$\pi$ or $\pi/2$.  As a result the asymptotic patterns that develop
can have four phases, two phases, or a mixture of two and four phases.

In this paper we explore pattern formation in the 4:1 resonance
regimes.  In Section~\ref{sec:experimental}, we describe our
experimental observations of four phase patterns in the 4:1 resonance
band of the forced Belousov-Zhabotinsky reaction.  We then present an
analytical study of the 4:1 resonance~\cite{EHM:98,EHM:99} in
Section~\ref{sec:fourphase}.  The study is based on a normal form, or
amplitude equation, approach which is strictly valid only close to the
Hopf bifurcation of the unforced oscillatory system. In order to test
the analytical predictions and to study the behavior of forced systems
far from the Hopf bifurcation, which is the case in the experiments,
we conduct numerical studies of two reaction-diffusion models (the
FitzHugh-Nagumo and Brusselator).  We describe the models and results
in Section~\ref{sec:numerical}.  In Section~\ref{sec:conclusion} we
discuss and compare the analytical and numerical results with the
experimental observations.

\section{The periodically forced Belousov-Zhabotinsky reaction}
\label{sec:experimental}

We use a light-sensitive form of the Belousov-Zhabotinsky (BZ)
reaction, a chemical reaction system with oscillatory kinetics, to
study the 4:1 subharmonic resonance patterns. 
In the experiments, the chemicals of the BZ system 
diffuse and react within a $0.4$~$\rm mm$ thick porous
membrane.  The system is maintained in a non-equilibrium
steady state by a continuous flow of fresh, well mixed reactant
solutions~\cite{BZ}
 on either side of the thin membrane where the patterns form.  
The unforced pattern is a rotating spiral wave 
wave of ruthenium catalyst concentration.  

We periodically force the system using spatially homogeneous square
wave pulses of light with intensity $I$, where $I$ is the square of
the forcing amplitude, and pulse frequency $\omega_f$ ($\omega_f/2\pi$
in Hz).  We choose the frequency $\omega_f$ to be approximately four
times the natural frequency of the unforced oscillations.

\begin{figure}
\centering\includegraphics[width=3.25in]{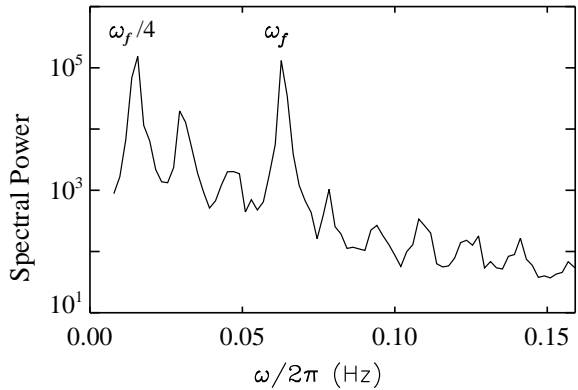}
\caption{
The temporal power spectrum of a 4:1 resonant pattern
from the BZ experiment for $I=426~\rm W/m^2$.  
The peak at $\omega/2\pi=0.0154$ Hz is the 
response at $\omega_f/4$.
}
\label{fig:bz_power}
\end{figure}

To determine the temporal response of a pattern 
when it is periodically perturbed at a particular pair
of ($I$,$\omega_f$) parameter values we collect a time series of
evenly sampled pattern snapshots; a $60\times60$ pixel region of
the $640\times480$ pixel image.
We sample at a rate of
approximately 30 frames/oscillation and calculate the Fast Fourier
Transform for the time series of each pixel.  The power spectrum of
each pixel is determined. An average over all pixels 
provides a power spectrum of a pattern, as shown in
Fig.~\ref{fig:bz_power}.  The 4:1 resonant patterns exhibit a dominant
peak at $\omega_f/4$ in the power spectrum.  Higher order harmonics
are also present.

\begin{figure}
\centering\includegraphics[width=3.25in]{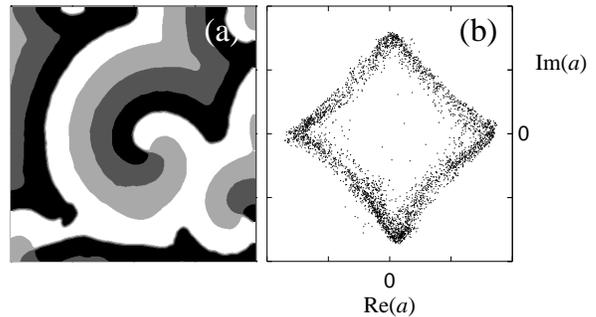}
\caption{
A rotating four-phase spiral wave observed in the forced BZ reaction.
(a) A $5.4\times5.4~\rm mm^2$ region of a reactor image showing a 4:1
resonant spiral pattern.  The white, light gray, dark gray and black
domains represent the four phase states of the system.  (b) A plot of
the complex Fourier amplitude $a$ at $\omega_f/4$ for each pixel of
the pattern.  The forcing intensity is $I$= 426
$\mbox{W}$/$\mbox{m}^2$ and the forcing frequency is $\omega_f/2\pi=
0.062$~Hz.  The data were filtered to isolate the response at
$\omega_f/4$ from the higher harmonics.
}
\label{fig:bz}
\end{figure}

An example of a 4:1 resonant pattern observed in the experiments is
shown in Fig.~\ref{fig:bz}.  The  rotating four-phase
spiral wave  in Fig.~\ref{fig:bz}(a) is the asymptotic state
of the system.  This image is a plot of the phase angle
$\arg(a)$, where $a=a(x,y)$ is the complex Fourier 
amplitude associated with the $\omega_f/4$ mode for each pixel $(x,y)$ in 
the pattern.  The four domains (white, light gray, dark
gray and black) correspond to the four phase states with oscillation phases
that are shifted by $0$, $\pi/2$, $\pi$, and $3\pi/2$  with respect to
the forcing.  

\begin{figure}
\centering\includegraphics[width=3.25in]{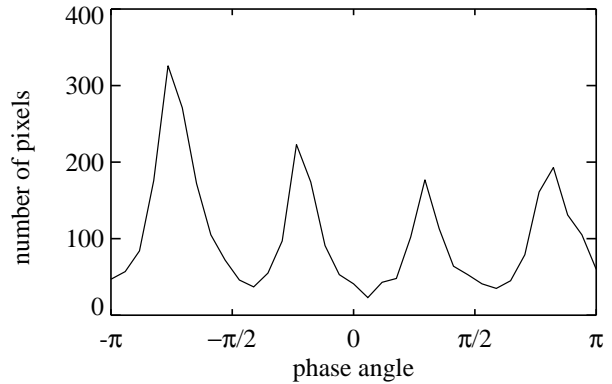}
\caption{ 
  A histogram showing the distribution
  of phase angles in the pattern in Fig.\protect~\ref{fig:bz}.  
  The four peaks indicate the high
  density of points in each of the four phase states.  
}
\label{fig:bz_histogram}
\end{figure}

Fig.~\ref{fig:bz}(b) is a different representation of the same data.
In this case the response $a$ at $\omega_f/4$ is plotted in the
complex plane instead of the $x-y$ plane.  This representation of the
data allows us to see the distribution of the oscillation amplitude
and phase at all pixels in the pattern.  
The four corners of the
diamond shape in Fig.~\ref{fig:bz}(b) are the four stable phase
states.  The edges of the diamond shape in Fig.~\ref{fig:bz}(b) are
formed from pixels at phase-fronts separating adjacent domains.  The
majority of pixels in the pattern are in one of the four corner states
as the histogram of phase angles in Fig.~\ref{fig:bz_histogram}
illustrates.

Traveling four-phase patterns exist over the entire dynamic range of 
forcing intensity $I$ in the 4:1 resonance region.  The range
of forcing intensity is limited by a $I$-dependence of the 
reaction kinetics.  As $I$ is increased, the reaction
kinetics shifts from oscillatory to excitable.

\section{An amplitude equation for forced oscillatory systems}
\label{sec:fourphase}

We study the experimental observations shown in the previous section 
using a normal form equation for the amplitude of the $\omega_f/4$ mode.
Consider first an oscillatory system responding to the forcing at 
$\omega_f/n$ where $n$ is integer. We assume the system is near the onset of
oscillations, i.e. close to a Hopf bifurcation.  The set of dynamical fields 
${\bf u}$ describing the spatio-temporal state of the system can be written as
\begin{equation}
{\bf u}={\bf u_{0}}A\exp {(i\omega_{f}t/n)} + c.c. + \dots,
\label{field}
\end{equation}
where ${\bf u_{0}}$ is constant, $A$ is a slowly varying complex
amplitude, and the ellipses denote other resonances with smaller
contributions.  The slow space and time evolution of the amplitude $A$
is described by the forced complex Ginzburg-Landau equation,
\begin{eqnarray}
A_\tau &=&(\mu +i\nu)A+(1+i\alpha )A_{zz}-(1-i\beta )|A|^{2}A
\nonumber \\ &&\mbox{}+\gamma_{n}{A^{\ast }}^{(n-1)}\,,
\label{fcglA}
\end{eqnarray}
where $\mu$ is the distance from the Hopf bifurcation, $\nu$ is the
detuning from the exact resonance, and $\gamma_n$ is the forcing
amplitude.

For the special case $n=4$ (the 4:1 resonance) we can eliminate the
parameter $\mu$ by rescaling time, space, and amplitude as
$t=\mu\tau$, $x=\sqrt{\mu/2}\,z$ and $B = A\sqrt{\mu}$ to obtain
\begin{eqnarray}
B_t &=&(1 +i\nu_0)B+\frac{1}{2}(1+i\alpha )B_{xx}-(1-i\beta )|B|^{2}B
\nonumber \\ &&\mbox{}+\gamma{B^{\ast}}^{3}\,,
 \label{fcglB}
\end{eqnarray}
where $\nu_0=\nu/\mu$.  Equation~(\ref{fcglB}) also applies to the 4:3
subharmonic resonance.  This follows from symmetry considerations: the
system is symmetric to discrete time translations $t\to
t+\frac{2\pi}{\omega_f}=t+\frac{3\pi}{2\omega}$. The amplitude
equation must then be invariant under the transformation $B\to
B\exp(3\pi i/2)$. The only forcing term satisfying this requirement to
cubic order is ${B^*}^3$.

\subsection{Phase states and phase fronts}

\begin{figure}[htb]
\centering\includegraphics[width=2.5in]{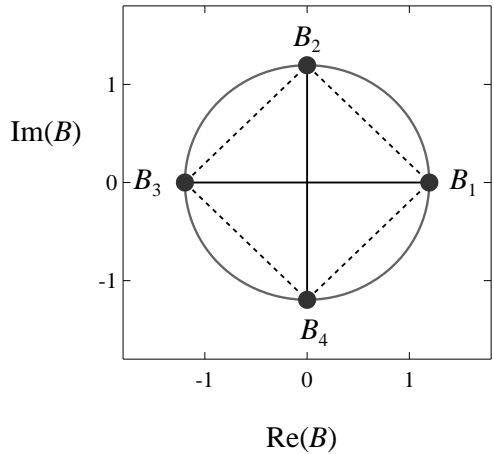}
\caption{
  The four phase states (black dots) connected by phase-fronts
  in the forced complex Ginzburg-Landau equation~(\protect\ref{gl}).  
  Two types of
  fronts between phase states are possible;  the solid lines are
  $\pi$-fronts and the dashed lines are $\pi/2$-fronts. 
}
\label{4:1-phase-none}
\end{figure}

Constant solutions of Eq.~(\ref{fcglB}) indicate that the system is entrained 
to the forcing. There are four stable constant solutions to Eq.~(\ref{fcglB}),
each with the same amplitude but with different 
phases, $\arg(B)$, which correspond to the four stable phase states. Simple 
expressions for these solutions and exact forms for the front
solutions connecting them in space are obtained from the
gradient version of Eq.~(\ref{fcglB}), where 
$\nu_0=\alpha=\beta=0$: 
\begin{equation}
B_t= B+\frac{1}{2}B_{xx}-|B|^2B +\gamma {B^{*}}^3\,.  \label{gl}
\end{equation}
The stable phase states (constant solutions) of Eq.~(\ref{gl}) for 
$0<\gamma<1$ are
$(B_1,B_2,B_3,B_4)=(\lambda, i\lambda, -\lambda, -i\lambda)$
where $\lambda=1/\sqrt{{1-\gamma}}$. They are 
represented as solid circles in Fig.~\ref{4:1-phase-none}.

Front solutions connecting pairs of these states are of two types,
fronts between states separated in phase by $\pi$ and fronts between
states separated in phase by $\pi/2$ (hereafter $\pi$-fronts and 
$\pi/2$-fronts).  The $\pi$-front solutions are
\begin{eqnarray}  \label{pifronts}
B_{3\to 1}&=&B_{1}\tanh{x}\,, \nonumber \\ B_{4\to
2}&=&B_{2}\tanh{x}\,.
\end{eqnarray}
For the particular parameter value $\gamma=1/3$ the $\pi/2$-fronts
have the simple forms
\begin{eqnarray}  \label{pi/2fronts}
B_{2\to
1}&=&\frac{1}{2}\sqrt{\frac{3}{2}}\bigl[1+i+(1-i)\tanh{x}\bigr]\,,
\nonumber \\ B_{1\to
4}&=&\frac{1}{2}\sqrt{\frac{3}{2}}\bigl[1-i-(1+i)\tanh{x}\bigr]\,,
\nonumber \\ B_{3\to 2}&=&-B_{1\to 4}\,, \nonumber \\ B_{4\to
3}&=&-B_{2\to 1}\,.
\end{eqnarray}
Additional front solutions follow from the invariance of
Eq.~(\ref{gl}) under reflection, $x\to -x$.

Figure~\ref{4:1-phase-none} shows these front solutions
(parametrized by the spatial coordinate $x$) in the complex $B$ plane. 
For example, the $\pi$-front $B_{3\rightarrow 1}$ is represented by the solid
line connecting the state $B_3$ to the state $B_1$ as $x$ increases from 
$-\infty$ to $+\infty$. The $\pi/2$-front $B_{2\to 1}$ is represented by the 
dashed line connecting the state $B_2$ to the state $B_1$.

In the special case of the gradient system~(\ref{gl}) all front
solutions are stationary.  The more general case with nongradient
terms in Eq.~(\ref{fcglB}) can be studied by
perturbation theory when $\nu_0,\alpha$ and $\beta$ are
small~\cite{EHM:99}.  The results of this analysis show that the
$\pi/2$-fronts become propagating fronts while the $\pi$-fronts remain
stationary.

\begin{figure}[htb]
\centering\includegraphics[width=3.25in]{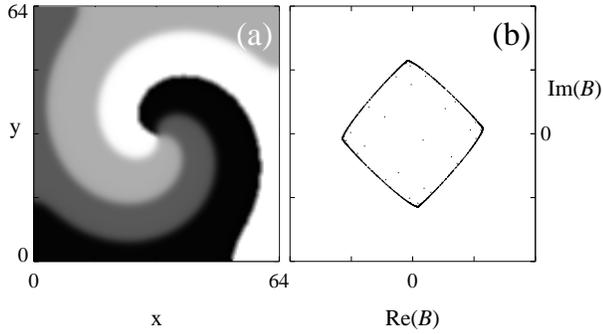}
\caption{
  A rotating four-phase spiral wave in the forced complex
  Ginzburg-Landau equation. 
  (a) $\arg(B(x,y))$ in the $x-y$ plane. 
  (b) $B(x,y)$ in the complex plane.  
  Parameters: $\gamma=0.6$, $\nu_0=0.1$, $\alpha=\beta=0$.
}
\label{fig:cglspiral}
\end{figure}

Figure~\ref{fig:cglspiral} shows a rotating four-phase spiral wave 
from a numerical solution of the two-dimensional version~\cite{cgl2}
of Eq.~(\ref{fcglB}). 
The phase diagram in the complex $B$ plane, shown in
Fig.~\ref{fig:cglspiral}(b), 
has four $\pi/2$-fronts: $B_{1\to 4}$, $B_{4\to 3}$,
$B_{3\to 2}$ and $B_{2\to 1}$.   
The amplitude $B$ corresponds to the complex Fourier amplitude $a$ 
measured in the experiment; the four-phase
spiral pattern in Fig.~\ref{fig:bz} and the corresponding diamond-shape 
in the complex plane are predicted by the amplitude equation.

\subsection{A phase-front instability}

The existence of the stationary $\pi$-front solutions 
suggests that standing two-phase patterns
similar to those found under 2:1 resonant conditions~\cite{POS:97,LBMS:99} 
may be observed in the 4:1 resonant case provided the $\pi$-fronts are stable.
Standing two-phase patterns have not been observed in experiments in the 4:1 
resonance band so the stability of $\pi$-fronts becomes a question.
Stability conditions for $\pi$-front solutions were studied in 
Refs.~\cite{EHM:98,EHM:99}. The results are described below.

\begin{figure}[htb]
\centering\includegraphics[width=3.0in]{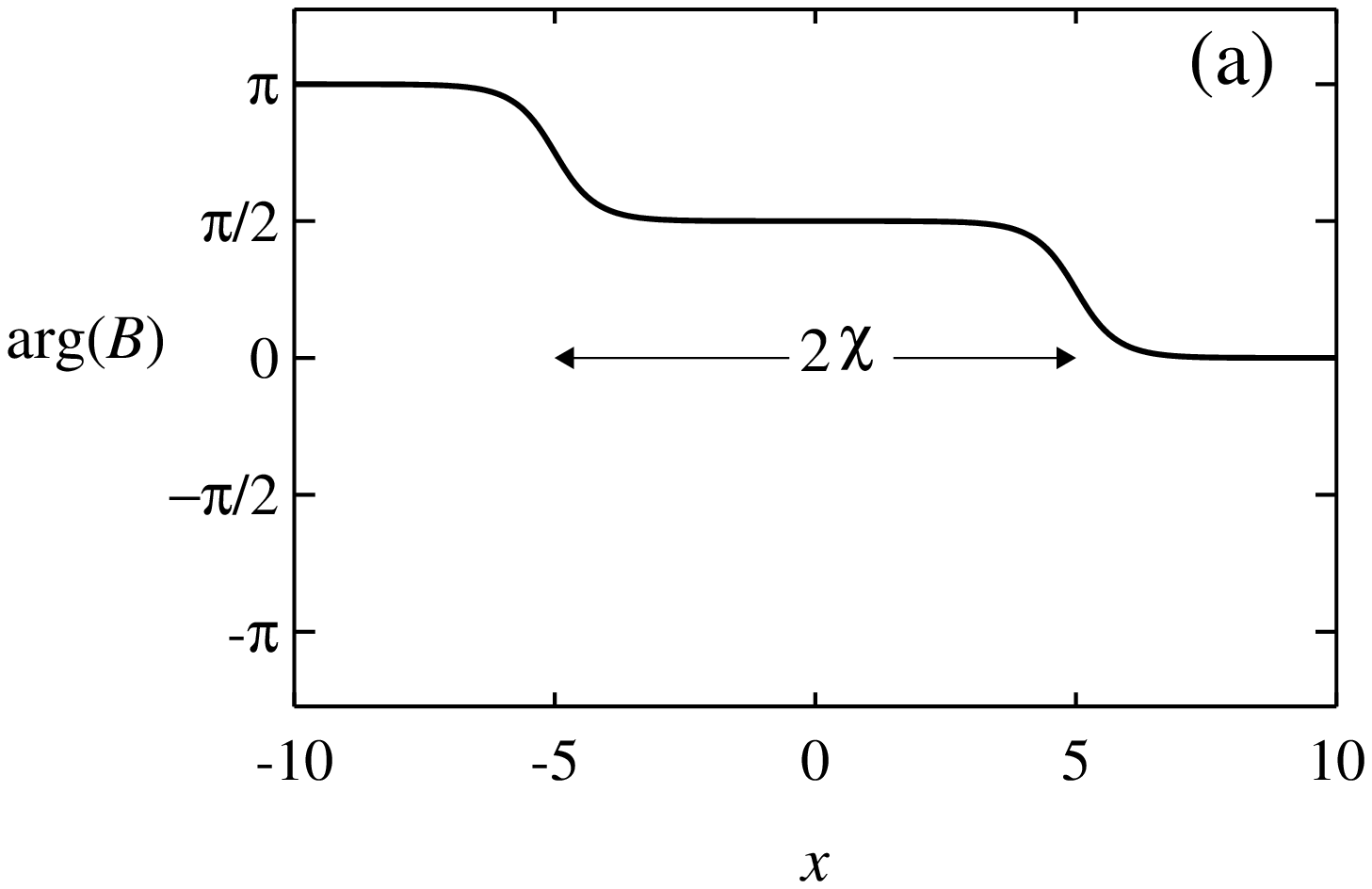}
\centering\includegraphics[width=3.0in]{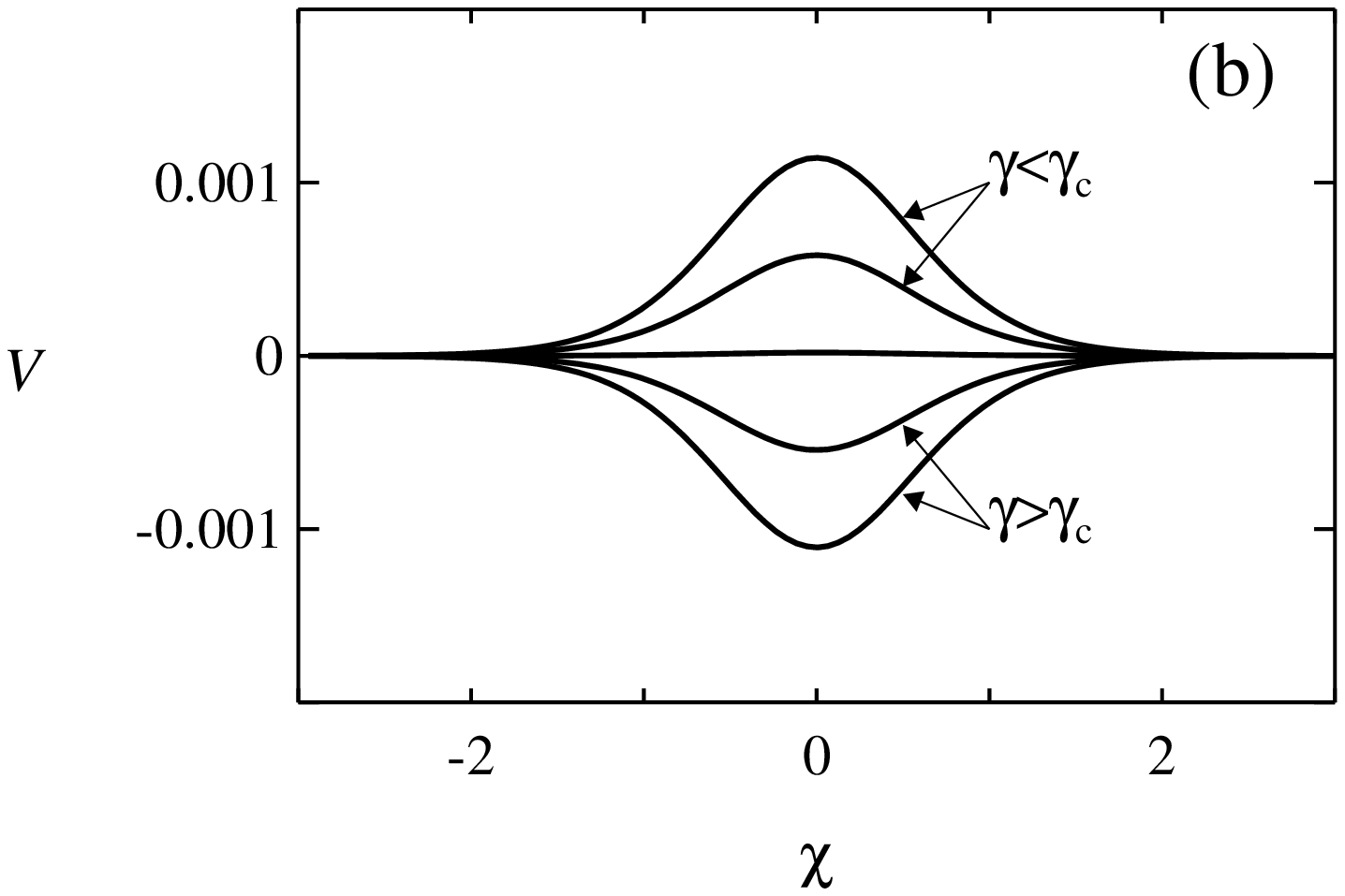}
\caption{
  (a) The phase, $\arg(B)$ of a pair of $\pi/2$-fronts,
   $B_{3\rightarrow 2}$, and $B_{2\rightarrow 1}$.  The distance
   between the two fronts is defined to be $2\chi$.  
   (b) The potential
   $V(\chi)$ describing the interaction between two $\pi/2$-fronts.
   For $\gamma>\gamma_c$ the extremum at $\chi=0$ is a minimum and
   $\chi$ converges to $0$.  For $\gamma<\gamma_c$ the extremum is a
   maximum and $\chi$ diverges to $\pm\infty$.  At $\gamma=\gamma_c$
   the potential is flat and there is no interaction between
   $\pi/2$-fronts.  }
\label{potential}
\end{figure}

Consider the pair of $\pi/2$-fronts shown in Fig.~\ref{potential}(a).
They are separated by a distance $2\chi$ and connect the phase states
$B_3$ and $B_1$.  For $\gamma \approx 1/3$, the
solutions~(\ref{pi/2fronts}) are good approximations to $\pi/2$-front
solutions. The pair of fronts can be represented as
\begin{equation}
B(x;\zeta,\chi)\approx B_{3\rightarrow
2}(x-\zeta+\chi)+B_{2\rightarrow 1}(x-\zeta-\chi)-i\lambda \,,
\label{eq:pairsol}
\end{equation}
where $\zeta$ is their mean position.  For large separation distances
($\chi>>1$) $B\approx B_{3\rightarrow 2}$ when $x\approx \zeta-\chi$
and $B\approx B_{2\rightarrow 1}$ when $x\approx \zeta+\chi$ and
Eq.~(\ref{eq:pairsol}) represents a pair of isolated $\pi/2$-fronts.
When the distance between the pair decreases to zero ($\chi\to 0$),
then $B\approx B_{3\rightarrow 1}$ and Eq.~(\ref{eq:pairsol})
approaches a $\pi$-front solution.

The stability of $\pi$-fronts is determined by the interaction between
a pair of $\pi/2$-fronts.  Stable $\pi$-fronts are the result of an
attractive $\pi/2$-front interaction; the $\pi/2$-fronts attract each
other and the distance between them decreases to zero. A repulsive
interaction implies unstable $\pi$-fronts.  The potential $V(\chi)$
that governs this interaction,
\begin{equation}
\dot{\chi}=-\frac{dV}{d\chi}\,,
\end{equation}
is shown in Fig.~\ref{potential}(b) for various $\gamma$ values.  The
potential has a single maximum for $\gamma< \gamma_c=1/3$ which
represents a repulsive interaction between $\pi/2$-fronts and the
instability of $\pi$-fronts.  It has a single minimum for
$\gamma>\gamma_c$ which indicates the attractive interaction between
$\pi/2$-fronts and the resulting stability of $\pi$-fronts.  At
$\gamma_c$ the potential is flat, $V=0$, for all $\chi$ values.  At
this parameter value, pairs of $\pi/2$-fronts do not interact and
there is a continuous family of front pair solutions with arbitrary
separation distances, $2\chi$, in Eq.~(\ref{eq:pairsol}).  This
degeneracy of solutions at the critical point $\gamma=\gamma_c$ is
removed by adding higher order terms to the amplitude equation, as we
discuss in Section~\ref{sec:higher}.

\begin{figure}[h]
\includegraphics[width=1.625in]{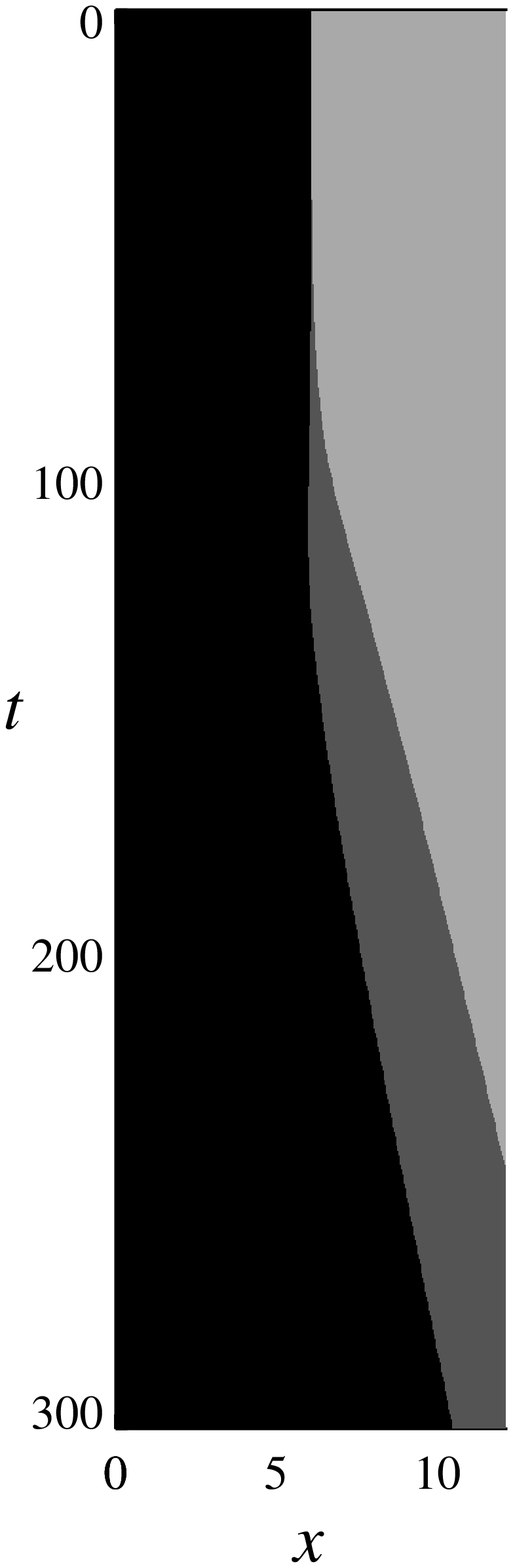}
\includegraphics[width=1.625in]{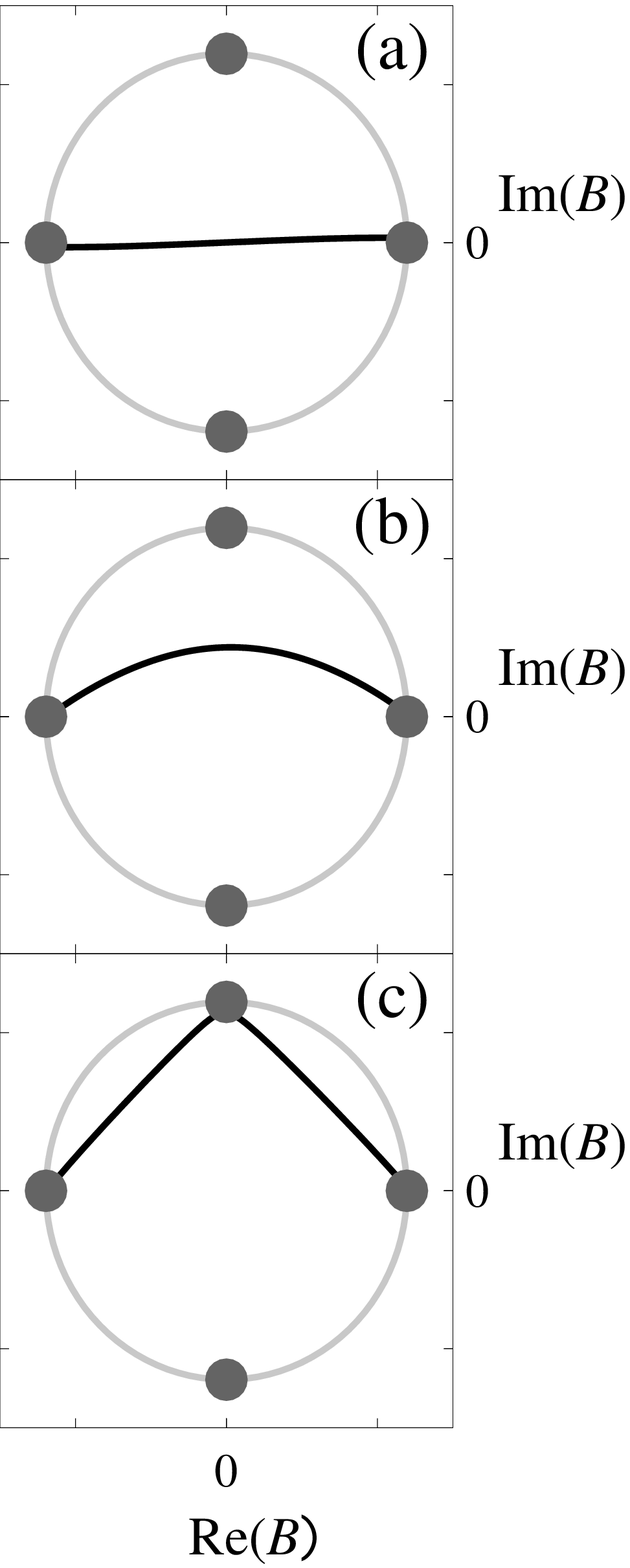}
\caption{
  An example of the phase-front instability in one space dimension.
  Left: The space-time plot (solutions of Eq.~(\protect\ref{fcglB}))
  shows the splitting of an unstable $\pi$-front into a pair of
  traveling $\pi/2$-fronts. The $\pi/2$-front pairs enclose the dark
  gray domain that has an oscillation phase shifted by $\pi/2$ with
  respect to the black and light gray domains.  Right: Snapshots at
  times $t=0$, $t=100$, and $t=300$, showing the instability in the complex
  $B$ plane.  Parameters in Eq.~(\protect\ref{fcglB}):
  $\nu_0=0.02$, $\gamma=0.3$, $\alpha=\beta=0$.
}
\label{fig:cgl-decomposition}
\end{figure}

To summarize, stationary $\pi$-front solutions of Eq.~(\ref{fcglB})
are stable for forcing amplitudes $\gamma>\gamma_c=1/3$.  When
$\gamma$ is decreased past $\gamma_c$, $\pi$-fronts lose stability and
split into pairs of propagating $\pi/2$-fronts. The splitting process
is shown in Fig.~\ref{fig:cgl-decomposition} where the $B_{3\to 1}$
$\pi$-front evolves into the pair of stable traveling $\pi/2$-fronts,
$B_{3\rightarrow 2}$ and $B_{2\rightarrow 1}$ when
$\gamma<\gamma_{c}$.  The parity symmetry $\chi\to -\chi$ makes
evolution toward the pair $B_{1\to 4}$ and $B_{4\to 3}$ equally
likely.  The splitting occurs for forcing amplitudes arbitrarily close
to $\gamma_c$, although in that case the time scale of this process
becomes very long.

\subsection{Effects of the phase-front instability on pattern formation}

The stability of stationary $\pi$-fronts for $\gamma>\gamma_{c}$
suggests the predominance of standing two-phase patterns.  These
patterns involve alternating domains with oscillation phases shifted
by $\pi$ with respect to one another. Domains shifted by $\pi/2$ may
exist as transients; the interactions between $\pi$-fronts and
$\pi/2$-fronts always produce $\pi/2$-fronts which are stable but
attract one another and coincide to form stationary
$\pi$-fronts. Since the $\pi/2$-fronts are traveling these transients
are relatively short.
For $\gamma < \gamma_{c}$ the interactions between the $\pi/2$-fronts
are repulsive.  The $\pi$-fronts are unstable and split into pairs
of traveling $\pi/2$-fronts.  As a result, traveling waves with
all four phase-states are the asymptotic pattern.  

\begin{figure}[htb]
\centering\includegraphics[width=3.0in]{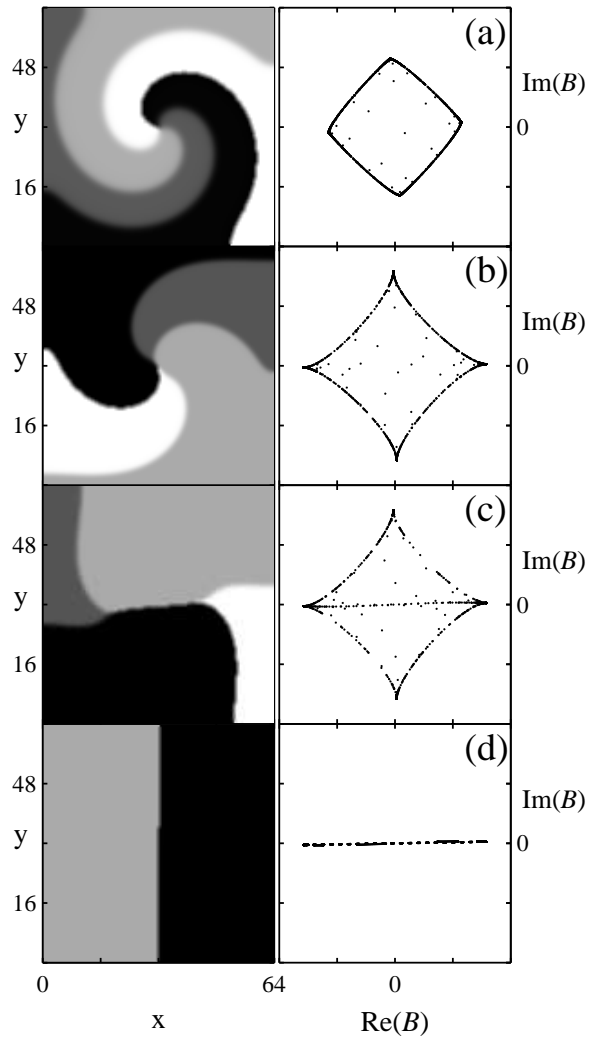}
\caption{
  Numerical solution of a two-dimensional version
  of Eq.~(\protect\ref{fcglB}) 
  showing the evolution of a rotating four-phase spiral-wave into
  a standing two-phase pattern when  $\gamma$ is increased above $\gamma_{c}$.
  The frames on the left show $\arg(B)$ in the $x-y$ plane.
  The frames on the right show the complex $B$ plane. 
  (a) The initial four-phase spiral wave
     (computed with $\gamma<\gamma_{c}$).  
  (b) The spiral core, a 4-point vertex, splits into two 3-point 
      vertices connected by a $\pi$-front.
  (c) A two-phase pattern develops as the 3-point vertices further separate.  
  (d) The final standing two-phase pattern.  
  Parameters: $\gamma=0.6$, $\nu_0=0.1$, $\alpha=\beta=0$, 
  $\gamma_c\approx 1/3$.
}
\label{fig:spiral}
\end{figure}

A typical two-dimensional traveling pattern involving all four phases
is the four-phase spiral wave shown in Fig.~\ref{fig:bz} or in
Fig.~\ref{fig:cglspiral}.  Figure~\ref{fig:spiral} shows the effect of
the phase-front instability on a four-phase spiral wave. The initial
spiral wave (Fig.~\ref{fig:spiral}(a)) was obtained by solving a
two-dimensional version of Eq.~(\ref{fcglB}) for
$\gamma<\gamma_{c}$. The following three frames
(Fig.~\ref{fig:spiral}(b)-(d)) are snapshots showing the evolution of
the initial four-phase spiral wave into a standing two-phase pattern
after $\gamma$ is increased above $\gamma_{c}$. The evolution begins
at the spiral core where the attractive interactions between pairs of
$\pi/2$-fronts are the strongest. The coalescence of $\pi/2$-fronts
leaves behind a stationary $\pi$-front which grows in length until no
$\pi/2$-fronts are left, as is evident by the single line in the
complex $B$ plane shown in Fig.~\ref{fig:spiral}(d).

\subsection{Higher order terms in the amplitude equation}
\label{sec:higher}

From the analysis of Eq.~(\ref{fcglB}) we have shown that two-phase
patterns must be standing and four-phase patterns must be traveling.
The analysis of the equation with higher order contributions suggests
the possible existence of a small $\gamma$ range, of order $\mu\ll 1$,
surrounding $\gamma_{c}$ where slowly traveling two-phase patterns
exist. 

\begin{figure}[htb]
\centering\includegraphics[width=3.0in]{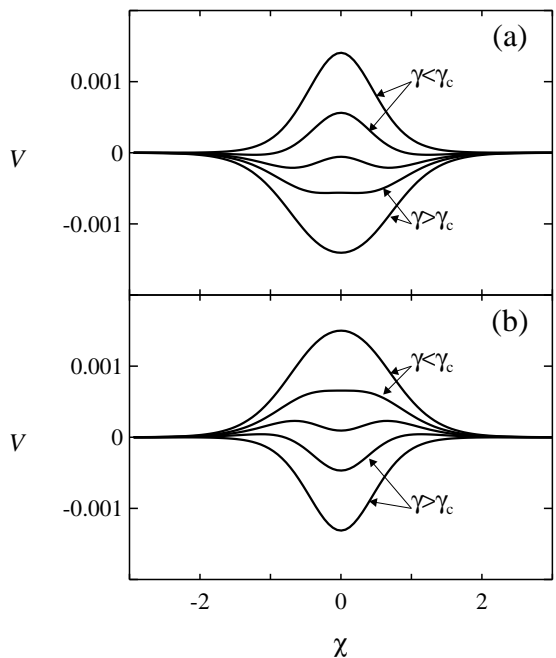}
\caption{
  The degeneracy of the potential $V(\chi)$ at $\gamma=\gamma_c$
  is broken by adding higher order terms to (\protect\ref{fcglB}).
  In the intermediate range of $\gamma\approx\gamma_c$ 
  two scenarios are possible as $\gamma$ is decreased through the bifurcation:
  (a)  The $\chi=0$ solution loses stability in a pitchfork bifurcation at
    $\gamma_{c}$ to a pair of solutions that move to $\pm\infty$.
  (b) The $\chi=0$ solution remains stable while the $\chi=\pm\infty$
  solutions acquire stability and lose stability only below
  $\gamma_{c}$.
  In both cases the deformations from a single minimum to a single
  maximum occur within a small range of $\protect\gamma$  of order
  $\mu\ll 1$.  }
\label{potential-higher}
\end{figure}

The higher order contributions to Eq.~(\ref{fcglB}), such as $|B|^4B$,
or $|B|^2 B_{xx}$, lift the degeneracy of the instability.
Figure~\ref{potential-higher} shows two possible scenarios for the
front interaction potential $V$ when higher order contributions to
Eq.~(\ref{fcglB}) are included (both scenarios lift the degeneracy of
the phase-front instability).  In one case, shown in
Fig.~\ref{potential-higher}(a), the stationary $\pi$-front loses
stability to a pair of counter-propagating $\pi$-fronts in a pitchfork
bifurcation which leads to double-minimum potential.  This scenario is
a nonequilibrium Ising-Bloch pitchfork bifurcation of $\pi$-fronts
like the one found in the 2:1 resonance case~\cite{CLHL:90} and in
other bistable systems~\cite{FRCG:94,IMN:89,HaMe:94a,BRSP:94}.  It
leads to slow traveling two-phase patterns in the range where $\gamma$
is near $\gamma_c$.  In the scenario shown in
Fig.~\ref{potential-higher}(b), the stationary $\pi$-front loses
stability via a subcritical bifurcation which leads to double-maximum
potential.  In this case there is a range of stable $\pi$-fronts
coexisting with pairs of separated $\pi/2$-fronts.  This allows the
possibility of patterns containing both $\pi$-fronts and
$\pi/2$-fronts.  Beyond this range the potential has a single maximum
and $\pi$-fronts split into pairs of $\pi/2$-fronts.  Both scenarios
persist over a range of $\gamma$ of order $\mu$, the distance from the
Hopf bifurcation.

\section{Numerical solutions of periodically forced reaction-diffusion models}
\label{sec:numerical}
The amplitude equation analysis predicts the existence of a
phase-front instability near the Hopf bifurcation and hints at
possible modifications of the instability as the distance from the Hopf
bifurcation is increased. Our objectives in this section are to test
the existence of the instability in reaction-diffusion
models and to use the models to examine how the instability is modified
far from the Hopf bifurcation.

\subsection{The FitzHugh-Nagumo model}
We study a periodically forced version of the FitzHugh-Nagumo
equations
\begin{eqnarray}
\label{eq:fhn}
u_t&=&u-(1+\Gamma\cos{\omega_ft})u^3-v+\nabla^2 u\,, \\
v_t&=&\epsilon(u-a_1v)+ \delta\nabla^2 v \,.  \nonumber
\end{eqnarray}
The unforced model is obtained by setting $\Gamma=0$.  The uniform
state $(u,v)=(0,0)$ undergoes a Hopf bifurcation as $\epsilon$ is
decreased past $\epsilon_c=1/a_1$. The Hopf frequency is
$\omega_H=\sqrt{\epsilon_c-1}$ and the distance from the Hopf
bifurcation is measured by $\mu=(\epsilon_c-\epsilon)/\epsilon_c$.

We compute the numerical solutions of Eq.~(\ref{eq:fhn}) in the 4:1
resonance band ($\omega_f\approx 4\omega_H$) and close to the Hopf
bifurcation ($\mu \ll 1$).  Close to the Hopf bifurcation the
amplitude equation analysis applies.  We expect to find a critical
value of the forcing amplitude $\Gamma_c$ corresponding to the
phase-front instability point $\gamma_c$ in the amplitude equation.
For the FitzHugh-Nagumo equations this $\Gamma_c$ will, in general,
depend on the parameters $\epsilon$, $\delta$, $a_1$, and $\omega_f$.
In the following we fix $a_1=1/2$, $\delta=0$, $\omega_f=4$ and only
vary $\epsilon$ (the parameter that controls the distance $\mu$ to the
Hopf bifurcation) and the forcing amplitude $\Gamma$.

\begin{figure}[htb]
\centering
\includegraphics[width=1.625in]{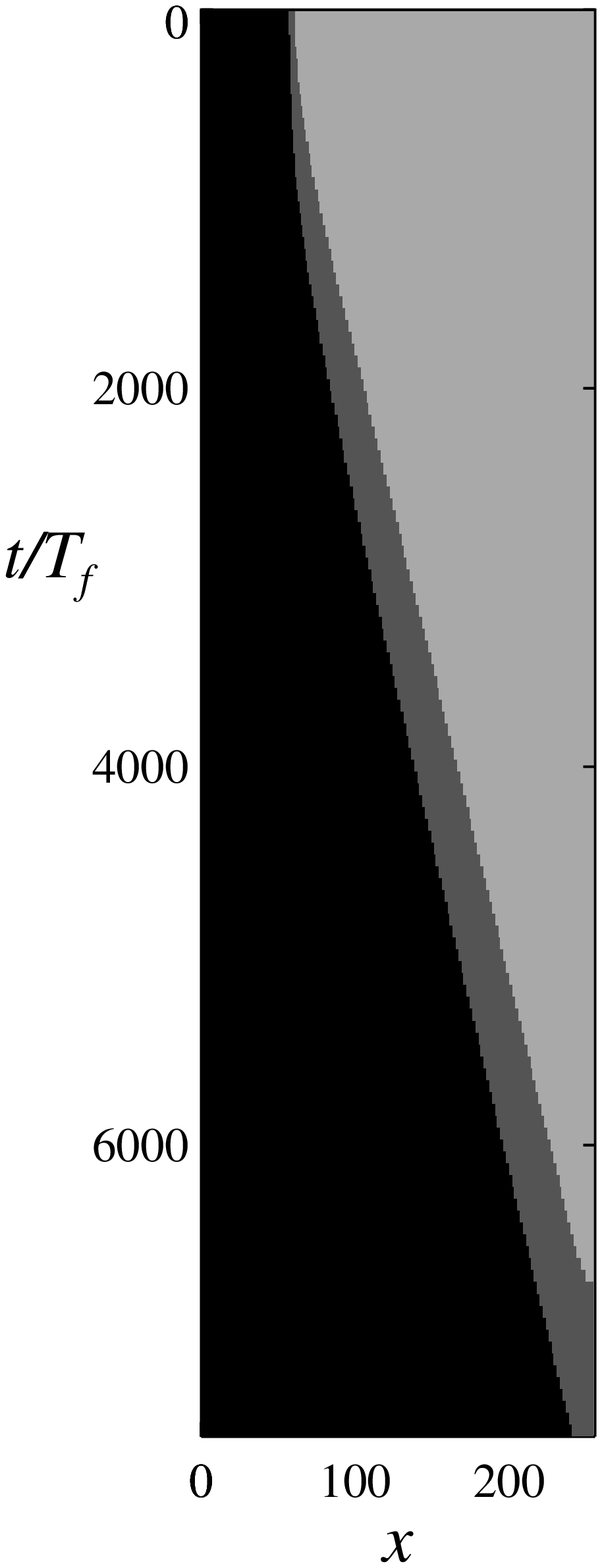}
\centering\includegraphics[width=1.625in]{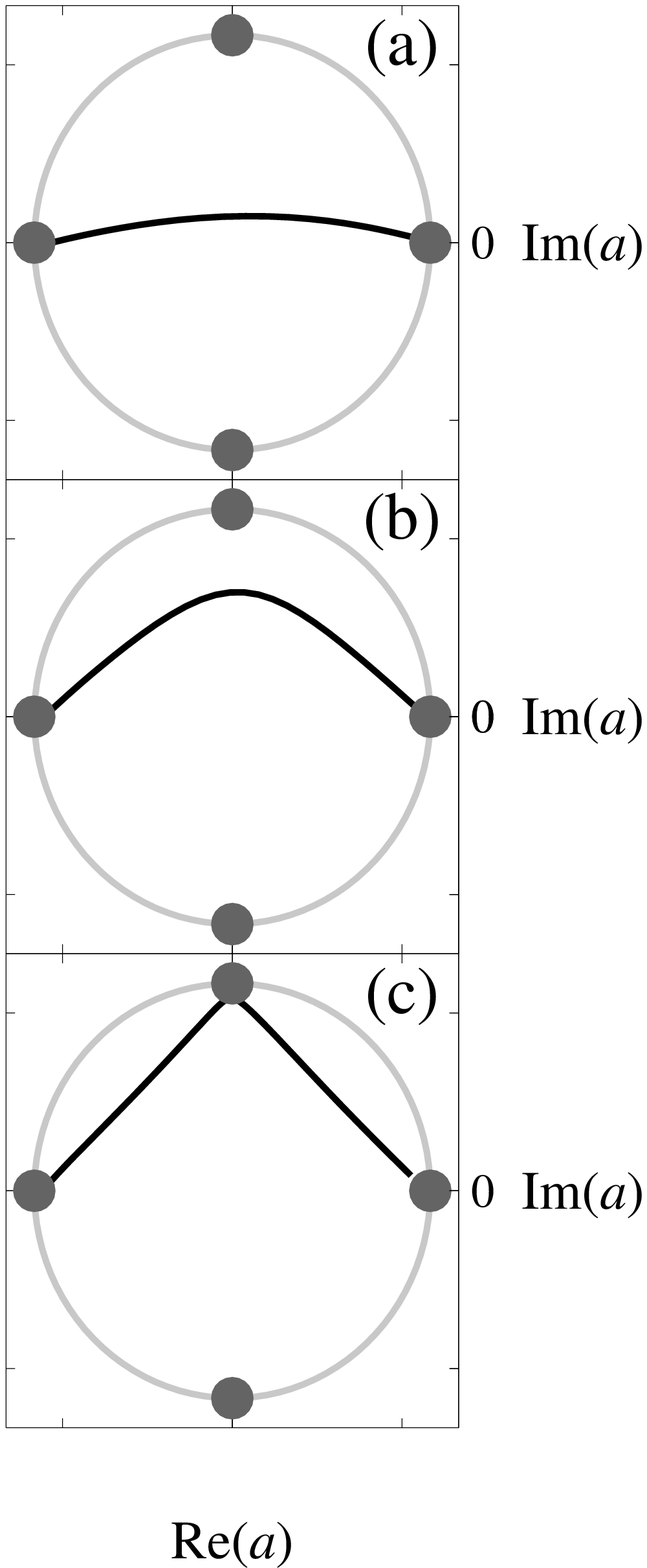}
\caption{
  The phase-front instability in
  the 4:1 resonance of the forced FitzHugh-Nagumo model
  close to the Hopf bifurcation. 
  Left: a space-time plot of $\arg(a)$ where $a$ is the 
  complex Fourier coefficient of the 4:1 response (equivalent to
  $A$ in Eq.~(\protect\ref{fcglA})).
  At $t=0$ the forcing amplitude was decreased below $\Gamma_c$.
  The initial standing  $\pi$-front becomes unstable
  and splits into a pair of traveling $\pi/2$-fronts.
  The $\pi/2$-fronts separate the black, dark gray, and light gray
  domains where the oscillation phase is shifted successively by $\pi/2$. 
  Right:  The same data depicted in the complex $a$ plane at three successive
  times, $t=0$, $t=560T_f$, and $t=4160T_f$ where $T_f=2\pi/\omega_f$.
  (a) The initial standing $\pi$-front is unstable.
  (b) The front develops an intermediate phase.
  (c) Two $\pi/2$ fronts are formed. 
  Parameters: $a_1=0.5$, $\epsilon=1.95$, $\delta=0$,  $\Gamma=2.0$,
  $\omega_f=4.0$, and $\mu=0.025$.  The phase-front instability point is
  $\Gamma_c\approx 2.15$ and $\eta\approx 0.012$.
}
\label{fig:fhn_near}
\end{figure}

Close to the Hopf bifurcation we find stable stationary $\pi$-fronts
for forcing amplitudes $\Gamma>\Gamma_c$.  Below $\Gamma_c\,$,
stationary $\pi$-fronts are unstable and split into pairs of
$\pi/2$-fronts.  Figure~\ref{fig:fhn_near} illustrates this in
numerical solution of a one-dimensional version of
Eq.~(\ref{eq:fhn}). An stable $\pi$-front pattern is generated from
random initial conditions with $\Gamma>\Gamma_c$.  At $t=0$ $\Gamma$
is decreased below $\Gamma_c\,$; the $\pi$-front becomes unstable and
splits into a pair of traveling $\pi/2$-fronts.

The numerically computed $\Gamma_c$ for the solution in
Fig.~\ref{fig:fhn_near} is $\Gamma_c\approx2.15$.  Since $\Gamma_c$ is
a function of the parameters in Eq.~(\ref{eq:fhn}), we define a new
parameter $\eta = (\Gamma _c-\Gamma)/\Gamma_c$ that measures the
distance from the phase-front instability point.  In
Fig.~\ref{fig:fhn_near}, $\eta \approx 0.012$ indicating that we are
just beyond the critical point.

\begin{figure}[htb]
\centering\includegraphics[width=1.625in]{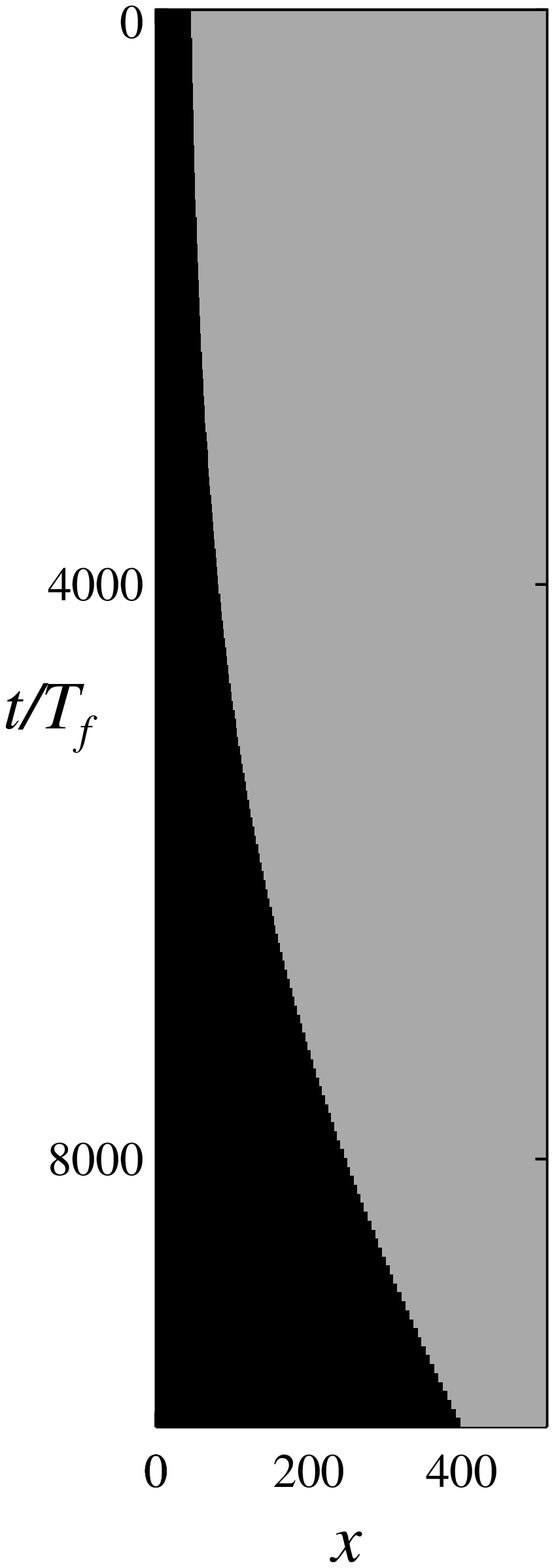}
\centering\includegraphics[width=1.625in]{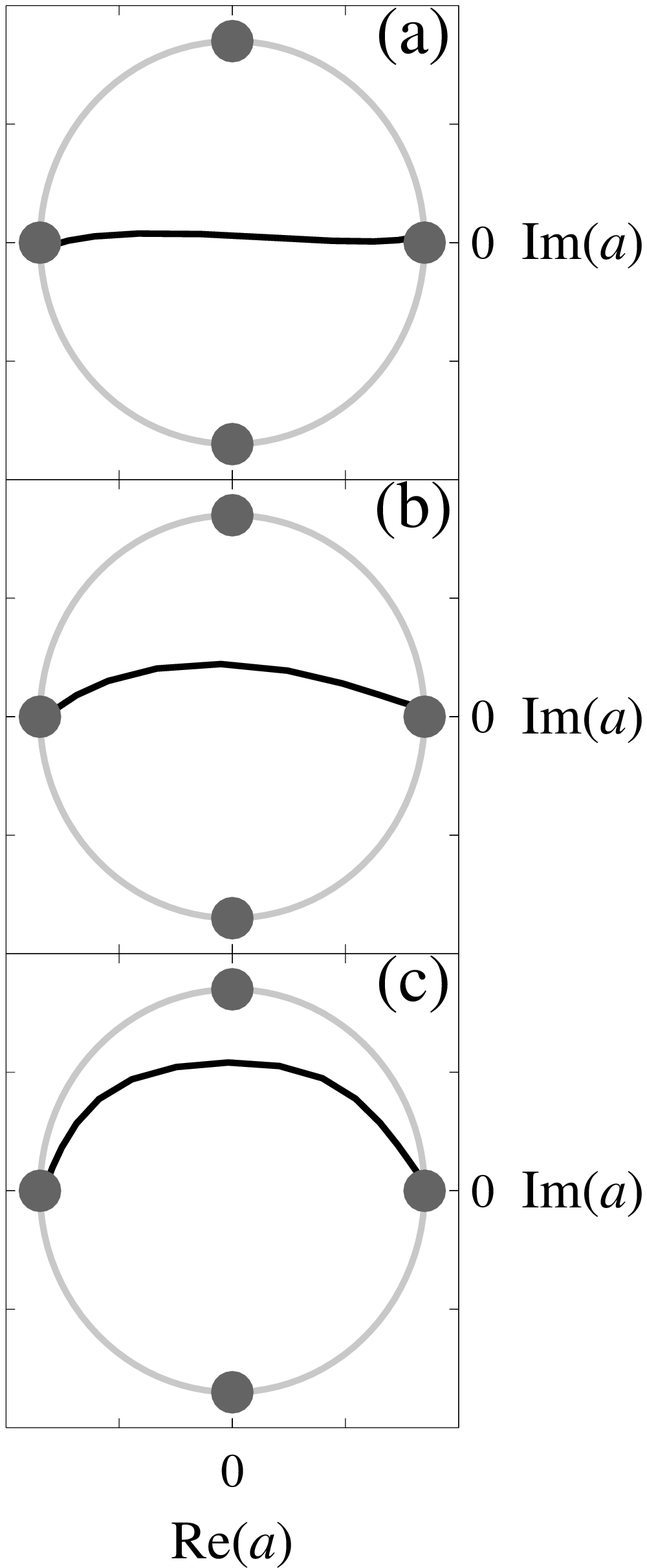}
\caption{
  The phase-front instability in the 4:1 resonance of the forced 
  FitzHugh-Nagumo model far from the Hopf bifurcation. 
  Left: a space-time plot of $\arg(a)$.
  At $t=0$ the forcing amplitude was decreased below $\Gamma_c$.
  The initial standing $\pi$-front is unstable and 
  and starts traveling to the right.
  In this case, no intermediate phase develops. 
  Right:  The same data depicted in the complex $a$ plane at three successive
  times, $t=0$, $t=3000T_f$, and $t=6000T_f$ where $T_f=2\pi/\omega_f$. 
  (a) The initial standing $\pi$-front.
  (b) The standing $\pi$-front is unstable and begins to travel.
  (c) The asymptotic pattern is a traveling  $\pi$-front.
  Parameters: $a_1=0.5$, $\epsilon=1.5$,  $\delta=0$,  $\Gamma=1.585$,
  $\omega_f=4.0$, and $\mu=0.25$.  
  The phase-front instability point is
  $\Gamma_c\approx 1.605$ and $\eta\approx 0.012$.
}
\label{fig:fhn_far_phase}
\end{figure}

Farther from the Hopf bifurcation we find that the phase-front
instability still exists.
Figure~\ref{fig:fhn_far_phase} shows the the evolution of an initial
unstable stationary $\pi$-front with parameters chosen so the system
is far from the Hopf bifurcation but at the same distance, $\eta
\approx 0.012$, from the phase-front instability.  The asymptotic
solution is a slowly propagating $\pi$-front, in contrast to a pair of
separated $\pi/2$-fronts that develop close to the Hopf bifurcation
(see Fig.~\ref{fig:fhn_near}).  The range of forcing amplitudes near
$\Gamma_c$ over which these traveling $\pi$-fronts exist increases
with $\mu$.  At smaller forcing amplitudes, below the range of
traveling $\pi$-fronts, $\pi$-fronts split into pairs of
$\pi/2$-fronts and four phase traveling patterns prevail.

\begin{figure}[htb]
\centering\includegraphics[width=3.0in]{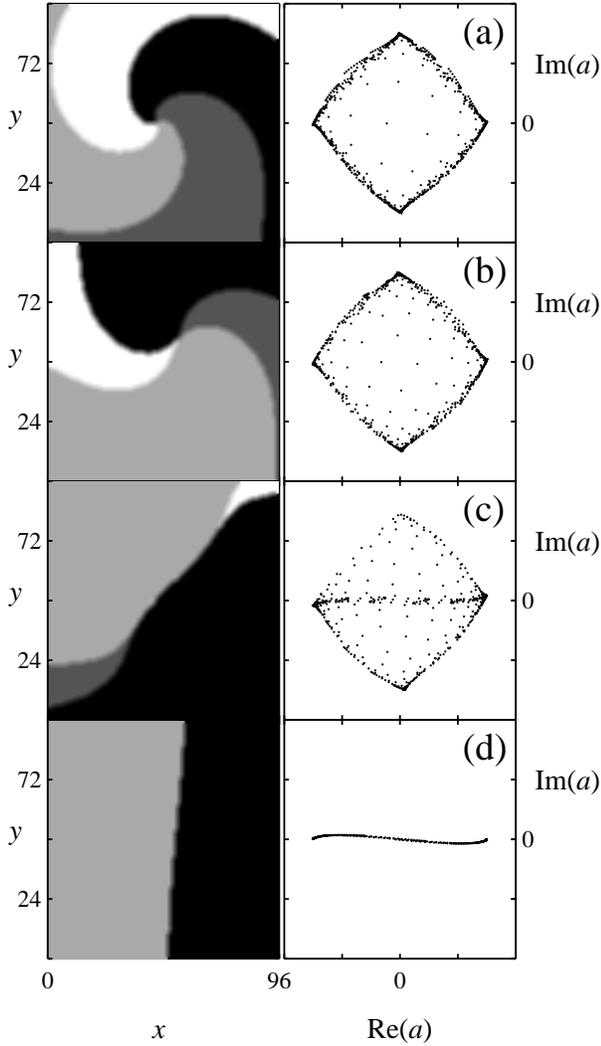}
\caption{
  Numerical solution of the forced
  FitzHugh-Nagumo equations~(\protect\ref{eq:fhn}) in 4:1 resonance
  shown at four successive times $t=0$, $t=11600T_f$, $t=13600T_f$, and
  $t=15600T_f$ where $T_f=2\pi/\omega_f$.
  The frames on the left show $\arg(a)$ in the $x-y$ plane.  The
  frames on the right show the complex $a$ plane.  
  (a) The initial
  spiral wave of four phases separated by $\pi/2$-fronts is computed
  with $\Gamma<\Gamma_{c}$.  
  (b) When $\Gamma$ is increased above
  $\Gamma_c$ two pairs of $\pi/2$-fronts begin to attract one another.  
  (c) As the $\pi/2$-fronts attract they collapse into
  a stationary $\pi$-front which grows in length.
  (d)  The final pattern is two phase domains separated by a stationary
  $\pi$-front.  Parameters: $a_1=0.5$, $\epsilon=1.5$, $\delta=0$,
  $\Gamma=2.5$, and $\omega_f=4.0$.
}
\label{fig:fhn_splitting}
\end{figure}

In two dimensions the typical traveling wave pattern for
$\Gamma<\Gamma_c$ is a rotating four-phase spiral wave.
Figure~\ref{fig:fhn_splitting}(a) shows a stable four-phase spiral wave 
generated from random initial conditions.
Using this spiral as 
an initial condition, we increase $\Gamma$ above
$\Gamma_c$ and the system evolves into a two-phase standing pattern.
Figures~\ref{fig:fhn_splitting}(b)-(d) show the transition.  Since the
$\pi/2$ fronts are attracting the spiral is unstable and two of the
four phase domains shrink until a standing two-phase pattern remains.

The numerical solutions of the forced FitzHugh-Nagumo equations
support the predictions of the amplitude equation analysis.  Close to
the Hopf bifurcation, the phase-front instability is found (compare
Fig.~\ref{fig:cgl-decomposition} with Fig.~\ref{fig:fhn_near} and
Fig.~\ref{fig:spiral} with Fig.~\ref{fig:fhn_splitting}).  Far from
the Hopf bifurcation the instability persists.  The effects of higher
order terms in the amplitude equation are valid even far from the Hopf
bifurcation ($\mu=0.25$); the phase-front instability near the Hopf
bifurcation (as $\mu\to 0$) turns into an Ising-Bloch pitchfork
bifurcation. Stationary $\pi$-fronts bifurcate to traveling
$\pi$-fronts and not $\pi/2$-fronts.

\subsection{The Brusselator model}

We tested the transition from four-phase traveling waves to two-phase
standing waves
using another reaction-diffusion model, the forced Brusselator,
\begin{eqnarray}
  \label{eq:brus}
  u_t&=&c-du+ [1+ \Gamma\cos{\omega_f t}]u^2v+\nabla^2 u\,, \\
  v_t&=&du-u^2v+\delta\nabla^2 v \,.  \nonumber
\end{eqnarray}
The unforced Brusselator, obtained by setting $\Gamma=0$, has a
stationary uniform state $(u,v)=(c,d/c)$ which undergoes a Hopf
bifurcation as $d$ is increased past $d_c=1+c^2$.  The Hopf frequency
is $\omega_H=c$ and the distance from the Hopf bifurcation is measured
by $\mu=(d-d_c)/d_c$.

\begin{figure}[htb]
\centering\includegraphics[width=3.0in]{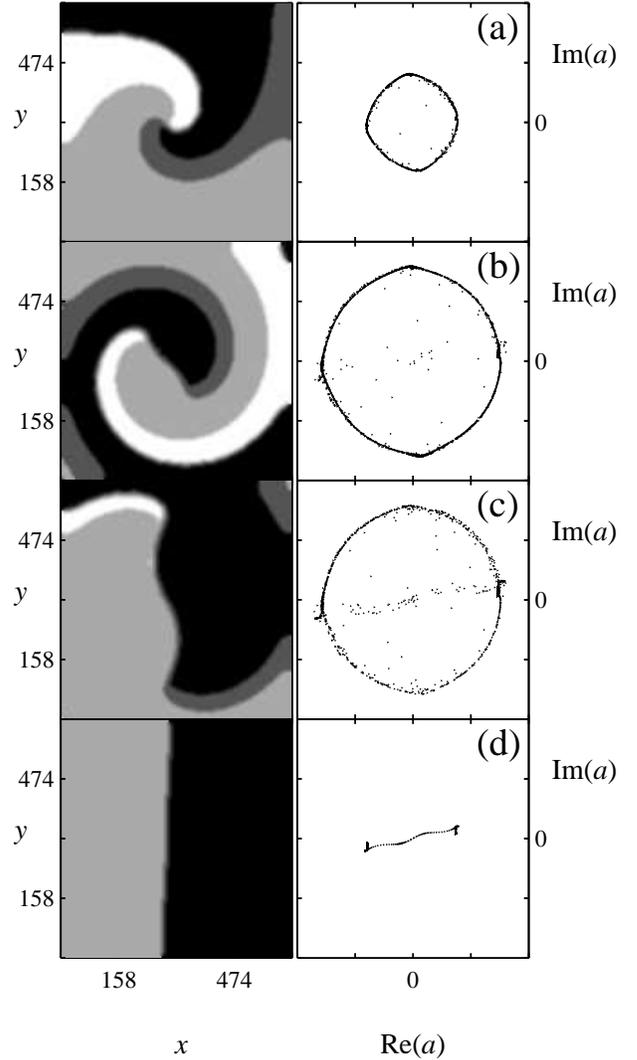}
\caption{
  Numerical solutions of the forced Brusselator 
  model~(\protect\ref{eq:brus}) showing snapshots at $t=0$,
  $t=748T_f$, $t=1000T_f$, and $t=5544T_f$ where $T_f=2\pi/\omega_f$.
  The  rotating four-phase spiral wave, computed with
  $\Gamma<\Gamma_c$ ($\Gamma=0.11$),  
  transforms into a standing two-phase pattern after $\Gamma$ 
  is increased past $\Gamma_c$ ($\Gamma=0.13$).
  The frames in the left column show $\arg(a)$ in the $x-y$ plane
  where $a$ is the  complex Fourier coefficient of the 4:1 mode.
  The right column shows the same data in the complex $a$ plane.
  Parameters: $c=0.5$, $d=1.5$, $\delta=5.0$, $\omega_f=1.69$, and
  $\mu=0.20$.  The numerical solution grid was $128\times128$ points.
  }
\label{fig:brusselator}
\end{figure}

We studied Eq.~(\ref{eq:brus}) in the 4:1 resonance band using a
numerical partial differential equation
solver~\cite{PaCa:96,APCMLS:00}.  We found that below a critical
forcing amplitude $\Gamma_c$ the solutions are rotating four-phase
spiral waves consisting of $\pi/2$-fronts (see
Fig.~\ref{fig:brusselator}(a)).  The four-phase spiral wave was
generated by one of two following initial conditions: a spiral wave
computed from the unforced ($\Gamma=0$) Brusselator equations, or the
linear functions
\begin{eqnarray*}
u(x,y)=& y/L,   & 0\le y \le L\,,\\ 
v(x,y)=&-2x/L+4 & 0\le x \le L \,,
\end{eqnarray*}
where $L=632.5$.

Above $\Gamma_c$ pairs of $\pi/2$ fronts attract each other and the core
of the spiral evolves into an expanding $\pi$-front.
Figures~\ref{fig:brusselator}(b)-(d) illustrate this process.  When
the $\pi/2$ fronts disappear, the resulting asymptotic pattern is two
states separated by a stationary $\pi$-front.
The transition from a four-phase spiral wave to a two-phase stationary
pattern, as in the amplitude equation model and the FitzHugh-Nagumo model, 
indicates the existence of the phase-front instability in the
Brusslator model.  

\section{Conclusions}
\label{sec:conclusion}

We studied 4:1 resonant patterns in Belousov-Zhabotinsky
chemical experiments, in an amplitude equation for forced oscillatory 
systems (the forced complex Ginzburg-Landau equation), and in forced 
FitzHugh-Nagumo and Brusselator reaction-diffusion models.  
At low forcing amplitudes all of these systems
exhibit traveling four-phase patterns.

An analysis of a forced complex Ginzburg-Landau equation, derivable
from periodically forced reaction-diffusion systems near a Hopf
bifurcation, predicts traveling four-phase patterns at low forcing
amplitude and standing two-phase patterns at high forcing amplitude.
The transition mechanism between these two patterns is a degenerate
phase-front instability where a stationary $\pi$-front splits into a
pair of traveling $\pi/2$-fronts.  We derived an interaction potential
between $\pi/2$-fronts that describes the instability as a change from
repulsive to attractive $\pi/2$-front interactions.  We investigated
the behavior of the instability near the critical point where higher
order terms in the amplitude equation become important.
We found that these terms lift the degeneracy of the instability and
introduce a narrow intermediate regime.  In this regime we found both
slowly traveling $\pi$-fronts and the coexistence of stable stationary
$\pi$-fronts and repelling pairs of $\pi/2$-fronts.

We further investigated this phase-front instability using the
FitzHugh-Nagumo and the Brusselator reaction-diffusion models.  These
models exhibit the instability even far from the Hopf bifurcation
where the amplitude equation is not known to be valid.  Near the Hopf
bifurcation the instability, at $\Gamma_{c}$, separates patterns of
stationary $\pi$-fronts from patterns of traveling $\pi/2$-fronts.  In
two dimensions, a rotating four-phase spiral wave evolves into a
two-phase standing pattern when $\Gamma$ is increased past $\Gamma_c$.
In the FitzHugh-Nagumo model we found, far from the Hopf bifurcation,
an intermediate range near $\Gamma_{c}$ where traveling $\pi$ front
patterns were observed.  These numerical results are in full agreement
with the theoretical predictions based on the amplitude equation.

The standing two-phase patterns found in the amplitude equation and in
the FitzHugh-Nagumo and Brusselator models were not observed in the
experiments, which were conducted far from the Hopf
bifurcation.  However, the existence of the phase-front instability
far from the Hopf bifurcation was found in the numerical studies of
the FitzHugh-Nagumo and Brusselator models.  We conclude that the
large distance from the Hopf bifurcation does not explain the absence
of standing two-phase patterns in the experiments.  A more likely
explanation is the limited dynamic range of the forcing amplitude in
the experiments.  Experiments show that the dynamics of the BZ
reaction are $\gamma$-dependent; as the forcing amplitude is
increased, the dynamics undergo a transition from oscillatory to
excitable kinetics.  The excitable kinetics are not described by the
amplitude equation or by the reaction-diffusion models in the
parameter ranges we studied.

\acknowledgments
We acknowledge the support of the Engineering Research Program of the
Office of Basic Energy Sciences of the U.S. Department of Energy.
Additional support was provided by the ASCI project B347883 through
the Lawrence Berkeley National Lab; the Robert A. Welch Foundation;
grant No. 98-00129 from the United States - Israel Binational Science
Foundation; and by the Department of Energy, under contract
W-7405-ENG-36.

\bibliography{fourphase}

\begin{thebibliography}{10}
\bibitem{BZ}
The chemical concentrations in the two chemical reserviors separated by the
  porous membrane were, Reservior I: $0.22$M malonic acid, $0.2$M sodium
  bromide, $0.264$M potassium bromate, $0.8$M sulfuric acid; Reservior II:
  $0.184$M potassium bromate, $1\times10^{-3}$M
  Tris(2,2'-bipyridyl)dichlororuthenium(II)hexahydrate, $0.8$ M sulfuric acid.
  Each reservior volume is 8.3 ml and the flow rate of chemicals through
  Reservior I was 20 ml/hr while through Reservior II it was 5 ml/hr. Chemicals
  were premixed before entering each reservior; a 10 ml premixer and a 0.5 ml
  premixer fed Reservior I and II, respectively. The experiments were conducted
  at room temperature.
\end{thebibliography}

\end{document}